\documentclass[12pt]{iopart}

\usepackage{graphicx}
\usepackage{dcolumn}
\usepackage{bm}
\usepackage{caption}
\usepackage{subfloat}
\usepackage{subfig}
\usepackage{float}
\usepackage{color}
\usepackage{setspace}
\usepackage{amssymb}
\usepackage{color,soul}
\usepackage{braket}

\begin{document}

 \title{General Formulation of Topos Many-Node Theory}

\author{Hamidreza Simchi}
\ead{simchi@alumni.iust.ac.ir}
\address{Department of Physics, Iran University of Science and Technology, Narmak, Tehran 16844, Iran}

\date{\today}

\noindent We consider the created entities (events) in the first moments of universe creation. It is assumed that there exists a causal energetic relationship between all events (nodes) such that all nodes are placed on a world line and each node occupies a region (instead of a point) in space-time, called locale, in mathematical terms. The set of locale nodes form a topos many-node system. Using some basic assumptions, we introduce two kinds of Hamiltonians. By attributing a general structural Hamiltonian to the system, it is shown that the system has an optimized critical dimension with a probable Raman and infrared spectrums. Also, we consider a general nonstructural Hamiltonian which includes a set of commutative self-adjoint operators and an interaction terms due to the spin, charge, or other kinds of probable degrees of freedoms for each $n^{th}$-optimized graph. For finding the state-space, truth values and quantity valued objects of the many-node system, a general procedure is introduced. The set of these values is a classical snapshot of the $n^{th}$-optimized graph which forms its kinematic. We show that the dynamic of the system can be explained by defining a combined map between the $n^{th}$- state-space belongs to the $n^{th}$-graph and the $({n+1)}^{th}$-state-space belong to $({n+1)}^{th}$-graph. Finally, by providing an interpretation of the general formulation of many-node theory, we discuss and explain how one can use the data of the cosmic background radiations and cosmic rays for finding a detailed model of both general structural and nonstructural introduced Hamiltonian. Here, time is no more than the change in truth value during comparison between $n^{th}$ and $({n+1)}^{th}$-graph.

\section{Introduction}
\noindent Newtonian physics including Newtonian mechanics, Maxwell equations, and statistical mechanics had three important conceptual achievements. The first one was the discovery of a procedure for describing macroscopic observable by using the equations governing the microscopic quantities and the second achievement was to introduce the concept of field in parallel to the concept of particle. The third one was the assumption of time fluid to explain the precedence and delay of events in nature and the acceptance of the existence of space (volume) as a place for events to occur.

\noindent There are three important conceptual achievements in relativity theory. It introduced the concept of space-time, the existence of the geometrical space-time fields as same footing as the existence of material fields and its geometricalization [1]. Of course, people like Weinberg [2] believed that the geometrical interpretation was not a necessity but a kind of taste in the interpretation of mathematical relationships to interpret some physical results. He believed without the use of geometricalization concepts, theoretical results could be compared with laboratory data and the adaptation or non -adaptation between them was properly explained [2].

\noindent The three important conceptual achievements of the theory of quantum physics (i.e., quantum mechanics and quantum field theory) were to introduce the concept of the state of a system instead of its position, velocity, and so on, the concept of the correlation between the state of a system and a complete experimental process including system preparation and measurement, and the concept of entanglement between twin entities. Therefore, in this attitude, we deal with the state of matter (whatever is not space-time fields) and its temporal evolution, and calculate the probability of an outcome of an experiment based on the information probability theory [3].

\noindent Then, we have been on a two -way for finding a convergence between the concepts of general relativity and the concepts of quantum physics at Planck scale: to quantize the theory of geometricalization of physic (relativistic based theories) [4 to 7], or to geometricalize the existing quantum theory (quantum fields based theories) [8 to 15]. But, a question can be asked: is it possible, we follow a third way for finding a convergence between the two sets of concepts i.e., whether can we follow a bottom-top approach for developing a new physical theory at Planck scale? But what do we mean by the bottom-top approach? For providing a better answer to the question, we pay our attention to the nanotechnology which has a very nice lesson for us. We can not only follow the top-bottom approach for finding nano-scale materials i.e., to exfoliate the bulk material for finding a single sheet of elements, but also bottom-top approach for manufacturing them by using the growing methods, such as the molecular beam epitaxy, for growing a single sheet of elements on a suitable substrate. Therefore, along to giving a positive answer to the above question, i.e., finding the convergence between quantum physics and general relativity at Planck scale, some people have tried to develop a new physical theory from beginning such as solid-state approach [16] and some background independent approaches [5, 17,18,19]. Table 1 shows some different old and new approaches which have been developed for understanding the physical world.

\noindent In this paper, we try to introduce a general formulation of topos many-node theory by using some key conceptual paradigms of causal set theory (CST) [20, 21], loop quantum gravity (LQG) [5, 22], topos quantum physics (TQP) [23-30] and quantum graphity [16, 31].  We borrow the concept of timid child in a causal partial ordered set from CST, N-node network and the concept of kinematic and dynamic from LQG, the concept of graph and its object and morphism from TQP and quantum graphity. Using the borrowed concepts and introducing some new assumptions, we consider a topos (Appendix A) as a graph including $N$-node and $N(N-1)/2$-link whose nodes form a causal sets which all of them are placed on a world line. The system is called many-node system. The nodes occupy a region in space-time, instead of a point, called locale in mathematical terms (Appendix B). For studying the system, we introduce two general types of Hamiltonian, called structural and nonstructural Hamiltonian. The general structural Hamiltonian will be used for studying the stability and surveillance of the graph. It is shown that, logically, there is a critical dimension for each optimized graph, which far from it for studying the system, we should use a new mathematical model and new conceptual paradigms [32]. Since, the final graph has an optimized (relaxed) structure, we can attribute the infrared and Raman spectrums and a maximized entropy [33] to the graph. Therefore, by using the data of cosmic background radiations and cosmic rays, it is possible to introduce a detailed model for each graph. The general nonstructural Hamiltonian will be used for finding the state-space, truth values, and quantity valued objects of the topos system. A general procedure is introduced for doing that (Appendix C). The set of these parameters stands for the kinematic of the system in each classical snapshot. We consider the evolution of the system as a collection of the classical snapshots i.e., it is shown that by defining a combined map, one can map the state-space of $n^{th}$-graph to the state-space of $({n+1)}^{th}$-graph. It is the definition of the system dynamics. Finally, we provide the interpretation of the theory including three important achievements. First, the locale nodes are used instead of points, and second, time is nothing more than the change in the physical parameters of events i.e., observed difference between the $n^{th}$-graph and the $({n+1)}^{th}$-graph. [32, 33]. Third one is the concept of probability. By using the concept of truth value, it is shown that in spite of the quantum theory, the probability is a derived concept.

\noindent The structure of the paper is a follows. A conceptual model, which is mostly used for developing a physical theory, and the necessary assumptions are provided in section 2. The kinematic and dynamic of the system are provided in section 3 and 4, respectively. Section 5 includes the interpretation and predictions of the resulted mathematical model. The summary is provided in section 6. As complementary materials, three Appendixes are provided. Appendix A provides the basic knowledge about the topos quantum mechanics. The concept of the locale, as a region in space-time, is provided in Appendix B. As an example, the topos representation of the 2-spin system is provided in Appendix C.
\section{Conceptual model and Assumptions}
\subsection{Conceptual model}

\noindent A review on the used applied methods for generalization or development of some theories often indicates that reference to new experimental data is the beginning of the theorizing process. But every scientist always has a set of basic mentality and past experiences, which may be used sometimes unconsciously. Also, she/he has the ability of self -awareness and analysis power to analyze the experimental data and to develop a mathematical model. The new model development can be done by generalization of the available models or even by creating new mathematics. Mathematical achievements alone are not sufficient to interpret natural events, and the philosophical interpretation of the relationships should be achieved for understanding the mathematical relationships, its compliance with pre-scientific achievements, as well as predicting possible future events. If the predictions of new developed theories are proven in labs, they would be accepted and the found results would be added to the previous possessions in the form of mentality and experience (Fig.1). If there are no available observations to begin theorizing, the theorizing process begins with mathematical modeling based on the previous experiences and possessions, but in the end, the philosophical interpretation, prediction of future events, and their proofs are necessary. The final theorization is subject to the approval of past experiences and future predictions. Due to the lack of technological and laboratory capabilities at present, the process of developing quantum gravity theory on the Planck scale begins from the mathematical model presentation. Therefore, in this article, we try to present a general mathematical model for kinematic and dynamic of the many-node theory which are necessary for providing a topo's interpretation of the theory.
\\
\\
\\
\\
\\
\\
\noindent \textbf{Table1- Some different approaches for understanding the physical world}
\begin{table}[h]
\centering
\tiny
\begin{tabular}{|p{0.6in}|p{0.5in}|p{0.1in}|p{0.1in}|p{0.4in}|p{0.4in}|p{0.6in}|p{0.7in}|p{0.7in}|p{0.7in}|} \hline
\textbf{Approach} & \multicolumn{2}{|p{0.6in}|}{\textbf{Space}} & \textbf{Time} & \textbf{Particle} & \textbf{Field} & \textbf{Sate of system} & \textbf{Kinematic} & \textbf{Dynamic} & \textbf{Gravity} \\ \hline
\textbf{Newtonian} & \multicolumn{2}{|p{0.6in}|}{volume} & fluid & exist & exist & Position, Velocity, Field strength and direction & Initial data & Evolution in time & Interaction between masses \\ \hline
\textbf{Relativity} & \multicolumn{3}{|p{1.0in}|}{Space-time} & exist & exist & Position, Velocity, Field strength and direction & Initial data & Evolution in layered space-time & Geometrical fields \\ \hline
\textbf{Quantum mechanics and Fields} & \multicolumn{3}{|p{1.0in}|}{Space-time or space and time} & exist & exist & Eigen functions & Hilbert space & Evolution in time or layered space-time & Negligible \\ \hline
\textbf{Loop Quantum Gravity} & \multicolumn{2}{|p{0.6in}|}{Quantized volume and area} & NA & exist & exist & Vertices, faces, edges & SL(2,C) and Diffeomorphism invariant Hilbert space & Wheeler--DeWitt equation & Spin Foam \\ \hline
\textbf{Topos Quantum Mechanics} & A region of universe (Locale) & \multicolumn{2}{|p{0.5in}|}{A combined map (Flow)} & An occupied region of universe & NA & Graphs,\newline objects and morphism & Category of abelian von Neumann sub-algebra & Snapshots of classical states & Topos representation\newline as locale and flow concepts \\ \hline
\textbf{Causal Set Theory} & \multicolumn{3}{|p{1.0in}|}{No happen in time but rather as constitute time} & exist & exist & Elements of partial ordered set & A partial ordered set & By adding an element step by step & Causal world line \\ \hline
\end{tabular}
\end{table}

\includegraphics*[width=6.46in, height=3.49in, keepaspectratio=false]{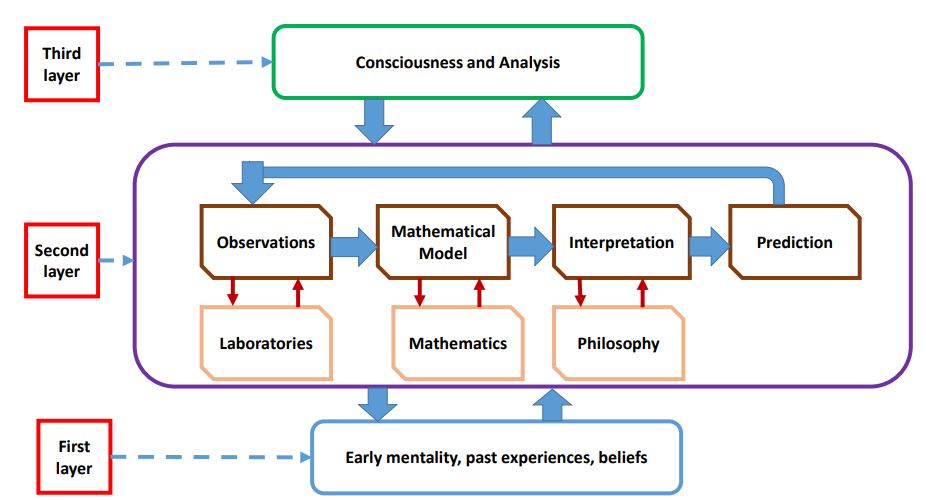}

\noindent

\noindent \textbf{Fig. 1 The conceptual model of scientific theorizing }
\subsection{Assumptions}

\noindent From causal set theory [20], we know that a causal set is a locally finite, partially ordered set (poset) which is denoted by C and a chain is a linearly ordered subset of C (a subset, every two elements of which are related by $\mathrm{\prec }$). A covering relations (called link) of a poset is an irreducible relation, and a path in a poset is an increasing sequence of elements, each related to the next by a link. Each element and each any two elements x $\mathrm{\prec }$ y related by a link are shown by dot and a line, respectively. The final graph is called the Hasse diagram. A causal set can be formed by accreting a single element to a second causal set (causet), then the former (the ``child'') follows the latter (the ``parent'') in a poset of finite causal sets and the relation between them is a link. The child formed by adjoining an element which is to the future of every element of the parent will be called the timid child [20]. The child formed by adjoining an element which is spacelike to every other element will be called the gregarious child [20]. The causal sets which can be formed by adjoining a single maximal element to a given causal set will be called collectively a family [20].

\noindent Our assumptions are as below:

\begin{enumerate}
\item  After adding an element to a causal set, if a causal relation is formed between the new element and the existed chain then the new element is placed on the world line of his parent. We call the new chain the ``timid child'' [33]. But, if by adding the new element a causal relation is not formed we call the new chain the ``gregarious child'' [33]. Therefore, as Fig.2 shows, the green dots (events) are placed on the world line of the first parent but the red dots are not. The causal relation is shown by blue line in the figure. Therefore, the graphs including the green (red) events show the timid (gregarious) child. Of course, you can also compare timid child and gregarious child in another way. A collection contains some numbers of graphs. If all the graphs in a collection are connected by links, that collection represents a timid child (green events in Fig.2). But, if there are graphs in the collection that are not connected to any graph, that collection shows a gregarious child (red events in Fig.2). There are two main differences between our model and Sorkin's model [20]. First, based on our assumption, a gregarious child never will be a parent of the next timid child related to the first parent, as shown in Fig.2, but in Sorkin's model it can be (see for example Figs. 2 and 3 of Ref.20). Second, in our model, each node occupies a region, instead of a point, in space-time which is called locale, in mathematical terms but in Sorkin's model it is considered as a point [20]. Here, locale nodes stand for space-time.

\item   The causal relation between events is energetic i.e., we deal with an energetic causal set [34, 35]. Therefore, by adding a new element to graph the amount of cohesive (internal) energy of the graph (which is a negative number) decreases by the quanta of energy and stored in it [33]. In consequence, the graph will be more stable than before by growing i.e., adding new element to the available timid child (graph). Of course, its growth continues till a critical dimension which will be explained more in the next section.

\item  Every single event (an element of a chain) stands for a fundamental entity which belongs to the first moment of the universe creation. The final timid child is called the final graph. It should be noted that there can exist many final graphs at the first moment of the universe creation which of course are not related to each other from causality relation point of view (i.e., no correlation exists between them) and in consequence they are known as distinguishable different non-related physical systems.

\item  We attribute to each single event of the final graph a set of the commutative self-adjoint operators which are related to the physical observable quantities.

\item  Time is not more than change in physical observable quantities of the distinguished physical system. If no change is seen at all, time is meaningless [32]. In the other words, we do not deal with the evolution of physical partial observables in time but with the correlations between $n^{th}$-timid child and ${(n+1)}^{th}$-child.\textbf{}
\end{enumerate}

\section{Kinematic}

\noindent Based on our above assumptions, a final timid child (graph) including $N$ nodes and $N(N-1)/2$ links is called a network. The system can be considered as a many-node (entity) system which can be analyzed by using two general Hamiltonians called nonstructural Hamiltonian denoted by $H^{Net}_T$ and structural Hamiltonian denoted by $H^L_T$. It should be noted that $H^{Net}_T$ is related to the physical observables and $H^L_T$ is related to the stability and survivability of the many-node system (i.e., network).

\noindent \includegraphics*[width=6.54in, height=3.37in, keepaspectratio=false]{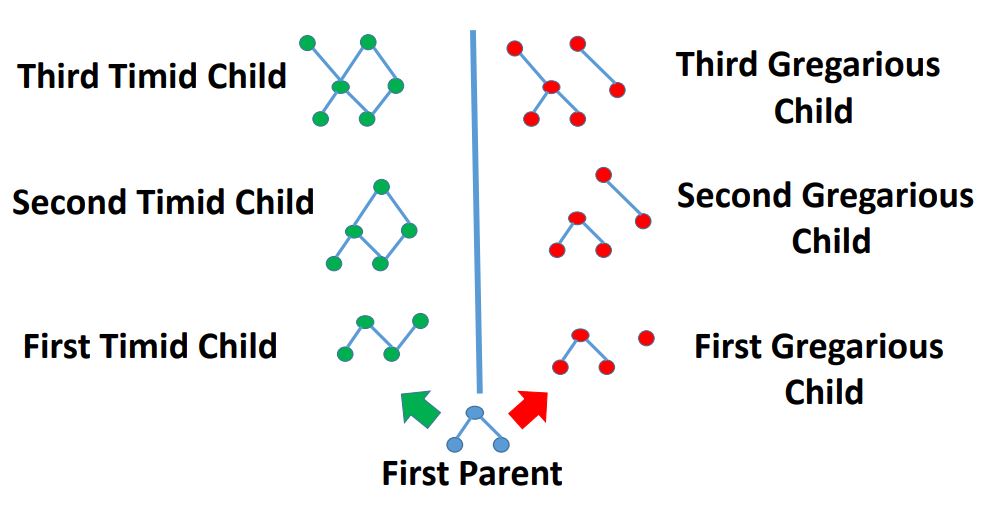}

\noindent \textbf{Fig. 2 An example of dynamic of a causal set. We only consider the timid children in our model. }

\noindent

\noindent Also, we use (without proof) a Born-Oppenheimer-like  approximation [37] and in consequence we separate the two general Hamiltonian $H^{Net}_T$ and $H^L_T$ from each other. The first question about the network is: whether is it an optimized (relaxed) network in the first moments of universe creation which both pressure and temperature are high? i.e., whether is it in its ground sate and stable in each classical snapshot?  It should be noted that based on the above assumptions, kinematic stands for one snapshot of correlated partial observables of a distinguished physical system. Here, the result of correlated partial observablea appear as the stability of the network in each snapshot. For studying the stability of the many-body system, let us assume $H^L_i$ stands for the general Hamiltonian related to the each fundamental entity in a locale node and $V^L_{i,j}$ stands for the general interaction potential between each fundamental entity and its neighbor fundamental entities (nodes) through links. Therefore, the total general structural  Hamiltonian in each snapshot may be written as $H^L_T$=$\sum^N_{i=1}{(H^L_i+\frac{1}{2}\sum^{M_i}_{j=1}{V^L_{i,j})}}$. Here, $L$ and $T$ stand for locale and total, respectively. $M_i$ is the number of linked neighbor nodes to each node, and the coefficient $\frac{1}{2}$ is inserted for removing the repeated terms. The balance between two terms $\sum^N_{i=1}{H^L_i}$ and $\frac{1}{2}\sum^N_{i=1}{\sum^{M_i}_{j=1}{V^L_{i,j}}}$ will guaranties the surveillance and stability of the network. Now, after stability of the final timid child, the total energy $H^L_T$ (called cohesive or binding energy) be a negative number, and is stored in the network. In the first stage of creation of a timid child, the cohesive energy becomes smaller and smaller by adding the new locale nodes (note that, it is a negative number) till the final timid child (graph) with a critical dimension is constructed. After reaching to the critical dimension, adding a new locale node increases the cohesive energy (i.e., it will be more positive than its previous stage) and the growth process stops. The critical step is a step that non-fundamental entities (events) start to created; timid child reaches to its critical dimension and we exit from the first moments of the creation of the universe. Of course, a new physical paradigm implies on the new physical system whose dimension is bigger than the critical dimension of the final timid child [32]. The calculation of the critical dimension of final timid child is an open question yet, which of course, should be answered by using the related experimental data. Logically, Raman and infrared spectrums are attributed to each relaxed final timid child structure (network). Therefore, it is natural to assume that there should be some spectrums related to the relaxed networks within the observed spectrum of the cosmic background radiations related to the first creation moments of the universe. These data may be used for finding the more exact theoretical model, and in consequence, the above mentioned critical dimension and other physical quantities of the final relaxed timid children. More explanations will be provided in section V.

\noindent But, what is about the nonstructural Hamiltonian $H^{Net}_T$ which is related to the physical observables such as spin, charge, parity, flavor, and so on? In topos representation of quantum theory, the system is considered as a locally snapshot of a classical system such that the collection of all snapshots includes all information about the system and nothing is lost (Appendix A). Since the physical quantities are real, a category of the snapshot, which is the category of the abelian von Neumann sub-algebra ($v(H)$) of the bounded operators on the Hilbert space ($B(H)$), should be considered [27, 28]. A standard procedure for providing topos representation of a theory is as follow (Appendix A):

\begin{enumerate}
\item  \textbf{First:} To specify the state-space (${\underline{\mathrm{\Sigma }}}_V$) and the sub-object classifier (${\underline{\mathrm{\Omega }}}_V$). The elements of state-space are the Gelfand spectrum of the self-adjoint operators $\hat{A}$ belongs to maximal algebra $V\in v\left(H\right)\ $and all sub-algebra $V^{\prime}\preccurlyeq V$. For doing that, first, the maximal algebra $V\in v\left(H\right)\ $and all sub-algebra $V^{\prime}\preccurlyeq V$ should be specified. Then the projection operators (${\hat{P}}_{\hat{A}}$) related to the self-adjoint operators $\hat{A}$ and their eigenvalues (${\lambda }_{\hat{P}}$) should be specified in $V$ and all $V^{\prime}s$. The set of all eigenvalues (${\lambda }_{\hat{P}}$) belong to $V$ ($V^{\prime}$) is the state-space ${\underline{\mathrm{\Sigma }}}_{V(V^{\prime})}$. Finally, the set of all sieves i.e., $\downarrow V=\left\{V^{\prime}\in v\left(H\right)\mathrel{\left|\vphantom{V^{\prime}\in v\left(H\right) V^{\prime}\preccurlyeq V}\right.\kern-\nulldelimiterspace}V^{\prime}\preccurlyeq V\right\}$ related to the maximal algebra $V$ and all $V^{\prime}s$, called  the sub-object classifier, should be specified which will be used for defining the truth values related to the physical quantities. (e.g, first step in Appendix C).

\item  \textbf{Second:} There are the cases in which the system is in the eigenfunction ($\ket{\psi }_A$) of a projection operator ${\hat{P}}_A$ related to the operator ${\hat{Q}}_A$ which belong to $A$-algebra but the quantity we want to measure is in the eigenfunction ($\ket{\psi }_B$) of the projection ${\hat{P}}_B$ related to the operator ${\hat{Q}}_B\ $which belong to $B$-algebra($\subseteq $ A-algebra), and is not an eigefunction of ${\hat{P}}_A$. For doing that, we should calculate the inner and outer deseinisations, i.e.,  ${\delta }^i{\left(\hat{P}\right)}_B$ and, ${\delta }^o{\left(\hat{P}\right)}_B\ $, respevtively. Then, by using the Eq. (A.1) one can find the approximated operators of ${\hat{Q}}_A$ in $B$-context (e.g, second step in Appendix C).

\item  \textbf{Third:} To specify the quantity valued object. For doing that we should use the above calculated approximated operators ${\delta }^i{\left(\hat{P}\right)}_B$ and, ${\delta }^o{\left(\hat{P}\right)}_B\ $of ${\hat{Q}}_A$ and find the values $a=\left\langle {\psi }_A\mathrel{\left|\vphantom{{\psi }_A {\delta }^O{\left({\hat{S}}_Z\right)}_B {\psi }_A}\right.\kern-\nulldelimiterspace}{\delta }^O{\left({\hat{S}}_Z\right)}_B\mathrel{\left|\vphantom{{\psi }_A {\delta }^O{\left({\hat{S}}_Z\right)}_B {\psi }_A}\right.\kern-\nulldelimiterspace}{\psi }_A\right\rangle $ and $b=\left\langle {\psi }_A\mathrel{\left|\vphantom{{\psi }_A {\delta }^O{\left({\hat{S}}_Z\right)}_B {\psi }_A}\right.\kern-\nulldelimiterspace}{\delta }^O{\left({\hat{S}}_Z\right)}_B\mathrel{\left|\vphantom{{\psi }_A {\delta }^O{\left({\hat{S}}_Z\right)}_B {\psi }_A}\right.\kern-\nulldelimiterspace}{\psi }_A\right\rangle $. Now, the quantity valued object $\check{\delta }(Q)$ related to ${\hat{Q}}_A$ in $B$-context, when the system is in the state $\ket{\psi }_A$ , is placed at the range [$a$, $b$] (e.g,  third step in Appendix C).
\end{enumerate}

\noindent It is well known that the basic arena wherein $N$-body  quantum mechanics takes place is the tensor product space, $H^N\equiv H^1_{Node}\otimes H^2_{Node}\dots \otimes H^N_{Node}$ of $N$ single-body Hilbert space [37]. By using the Density Functional Theory (DFT), instead of solving a N-body problem, we can solve one-body problem including an effective potentials composed by Hartree-Fock, $V_{HF}$ ,and exchange-correlation,$V_{XC}$ ,potentials [37]. But, a N-entity (node) system is a causal set (poset) and in consequence it can be considered as a topos [27]. Therefore, given each locale node $N_i$, we consider $p$ self-adjoint operator$A^k_{sa}$, $k=1,\dots ,p$ related to $p$ physical observables. Here, $sa$ stands for self-adjoint. Now, one can sort these operators in $m$ different classes (e.g  $m_1$, $m_2$, {\dots}, $m_k$) such that in each class, $m_k$, the operators commute with each other and belong to the same Hilbert space, $H^{{N_i,m}_k}_{sa}$ (similar to Hamiltonian operator, $H$ ,Casimir operator $J^2$ and $J_z$ in angular momentum theory). Since in the network, the links show the energetic causal relation between locale nodes i.e, they stand for the mutual interaction between locale nodes, one can consider the interaction potential $V^{N_i}_{l_j}$ between each locale node $N_i$ and its neighbor node which is connected to it by the link, $l_j$. Therefore, the total Hamiltonian of each node $N_i$, $H^{N_i}_T\ ,\ $when we consider the $m_k$- class representation of the self-adjoint operators, can be written as below
\begin{equation} \label{GrindEQ__1_}
H^{N_i}_T=H^{{N_i,m}_k}_{sa}+\sum^{M_i}_{j=1}{V^{N_i}_{l_j}}
\end{equation}
Therefore, in the $m_k$-class representation, the total Hamiltonian of the network will be
\begin{equation} \label{GrindEQ__2_}
H^{Net}_T=\sum^N_{i=1}{\left(H^{{N_i,m}_k}_{sa}+\frac{1}{2}\sum^{M_i}_{j=1}{V^{N_i}_{l_j}}\right)}
\end{equation}
where, $Net$ means network and the coefficient $\frac{1}{2}$ is inserted for omitting the repeated terms.

\noindent At first step let us, consider only the term $\sum^N_{i=1}{H^{N_i,m_k}_{sa}}$  in Eq. 2 and assume that the $m_k$-class representation includes the maximum self-adjoint operators which $\lambda $ stands for its related Gelfand spectrum. Following the first step of the above mentioned procedure, the topological space of all multiplicative linear functionals with norm one associated to each algebra $V\in v(H)$ can be considered as the state space ${\underline{\mathrm{\Sigma }}}_V$, i.e., [27, 28]:
\begin{equation} \label{GrindEQ__3_}
{\underline{\mathrm{\Sigma }}}_V:=\left\{\lambda :V\to C\mathrel{\left|\vphantom{\lambda :V\to C \lambda \left(\widehat{\boldsymbol{1}}\right)=1}\right.\kern-\nulldelimiterspace}\lambda \left(\widehat{\boldsymbol{1}}\right)=1\right\}
\end{equation}
where, $\widehat{\boldsymbol{1}}$ is unit operator. Therefore, by finding the maximal algebra $V$ and all sub-algebra $\downarrow V=\left\{V^{\prime}\in v\left(H\right)\mathrel{\left|\vphantom{V^{\prime}\in v\left(H\right) V^{\prime}\preccurlyeq V}\right.\kern-\nulldelimiterspace}V^{\prime}\preccurlyeq V\right\}$, one can find ${\underline{\mathrm{\Sigma }}}_V$ and ${\underline{\mathrm{\Sigma }}}_{V^{\prime}}$ for each $V^{\prime}\preccurlyeq V$. It should be noted that there exist a correspondence between the contravariant (${\underline{\mathrm{\Sigma }}}_V$) and the covariant (${\overline{\mathrm{\Sigma }}}_V$) representation of state space as $\overline{\mathrm{\Sigma }}(\downarrow C)$=$\underline{\mathrm{\Sigma }}(C)$ where $C\in v(H)$ [38]. Also, the set of all sieves i.e., $\downarrow V=\left\{V^{\prime}\in v\left(H\right)\mathrel{\left|\vphantom{V^{\prime}\in v\left(H\right) V^{\prime}\preccurlyeq V}\right.\kern-\nulldelimiterspace}V^{\prime}\preccurlyeq V\right\}$ related to the maximal algebra $V$ and all $V^{\prime}s$, i.e., the sub-object classifier can be found. Since, some operators (${\hat{Q}}_B$) do not belong to the $m_k$- class representation (such as $J_x$ and $J_y$ which do not belong to $\left(H,\ J^2,\ J_z\right)$-representation of angular momentum theory) we follow the second step of the above procedure and find their related inner and outer deseinisations, i.e.,  ${\delta }^i{\left(\hat{P}\right)}_B$ and, ${\delta }^o{\left(\hat{P}\right)}_B\ $, respectively. Finally, following the third step of the above mentioned procedure we find the quantity valued object $\check{\delta }(Q)$ related to ${\hat{Q}}_A$ in  $B$-context.

\noindent But what is about the interaction term $\frac{1}{2}\sum^N_{i=1}{\sum^{M_i}_{j=1}{V^{N_i}_{l_j}}}$  in Eq. 2? Konopka et al., have inspired the form of the fermionic-like and bosonic-like  interactions and introduced two interaction potentials $H_{exc}$ (exchanged correlation) and $H_{add}$ (addition and subtraction) in their quantum graphity model [31]. The term $H_{exc}$ preserves the valence of each vertex and the term $H_{add}$ adds or subtracts edges [31]. But in our model, we deal with a finial timid child including all locale entities (nodes) and all preserved links in each snapshot of the classical case. Therefore, we cannot follow a strategy similar to them. But, it is well known that in second quantization languages, the general form of the interaction potential between two neighbor nodes $A$ and $B$ might be written as  $\left\langle {\ \psi }^+_A,{\psi }^+_B\mathrel{\left|\vphantom{{\ \psi }^+_A,{\psi }^+_B V(A,B) {\psi }_B,{\psi }_A}\right.\kern-\nulldelimiterspace}V(A,B)\mathrel{\left|\vphantom{{\ \psi }^+_A,{\psi }^+_B V(A,B) {\psi }_B,{\psi }_A}\right.\kern-\nulldelimiterspace}{\psi }_B,{\psi }_A\right\rangle $ and in consequence the total interaction potential, $V_{Int}$, has the below form
\begin{equation} \label{GrindEQ__4_}
V_{Int}=\frac{1}{2}\sum_{A,B}{\ \left\langle {\ \psi }^+_A,{\psi }^+_B\mathrel{\left|\vphantom{{\ \psi }^+_A,{\psi }^+_B V(A,B) {\psi }_B,{\psi }_A}\right.\kern-\nulldelimiterspace}V(A,B)\mathrel{\left|\vphantom{{\ \psi }^+_A,{\psi }^+_B V(A,B) {\psi }_B,{\psi }_A}\right.\kern-\nulldelimiterspace}{\psi }_B,{\psi }_A\right\rangle \ }\
\end{equation}
Therefore, logically, we can assume that the interaction term $\frac{1}{2}\sum^N_{i=1}{\sum^{M_i}_{j=1}{V^{N_i}_{l_j}}}$ has the general form as shown by eq.(4). But, since we deal with the entities belong to the first moments of the universe creation, it includes different types of interactions such as $\sum_i{\sum_j{J_{ij}{\hat{S}}_{1i}{\hat{S}}_{2j}}}$ (spin-interaction), $\sum_i{\sum_j{Q_{ij}{\hat{I}}_{1i}{\hat{I}}_{2j}}}$ (charge-interaction) and son on. It means that $V_{Int}$:=$\ \frac{1}{2}\sum^N_{i=1}{\sum^M_{j=1}{V^{N_i}_{l_j}}}$ includes some terms such as ${\hat{S}}_{Ai}$, ${\hat{I}}_{Ai}$, and so on related to spin, charge, flavor and other probable interaction degree of freedoms. Now, if these operators belong to the chosen $m_k$-class representation, one can follow the first step of the above mentioned procedure and calculate $V_{Int}$, and if do not belong, she/he should follow the second step of the above mentioned procedure and then find the interaction potential $V_{Int}$ by using the approximated operators ${\delta }^i{\left(\hat{P}\right)}_B$ and, ${\delta }^o{\left(\hat{P}\right)}_B$.

\noindent Therefore, we can study the many-entity theory in each classical snapshot by using the above mentioned procedure and specifies the kinematic of the system. But, what is about the dynamic of the system. We will justify the subject in the next section.

\section{Dynamic}

\noindent Based on our above assumptions, we deal with the relationship between $n^{th}$-child and ${(n+1)}^{th}$-child. Therefore, dynamics means the relation between the different classical snapshots. In each snapshot, one can consider an unital $C^*$-algebra $A$ with associated category $C(A)$ of unital abelian $C^*$-subalgebra of $A$ which forms a poset under algebra inclusion such that the spectral presheaf ${\underline{\mathrm{\Sigma }}}^A_n$ belongs to ${\boldsymbol{Sets}}^{{C(A)}^{op}}_n$[28]. Here, $n$ is a subscript which is used for specifying the $n^{th}$-snapshot. Similarly, the spectral presheaf ${\underline{\mathrm{\Sigma }}}^B_{n+1}$ can be defined which belong to ${(n+1)}^{th}$-snapshot. If $\boldsymbol{u}{\boldsymbol{C}}^{\boldsymbol{*}}$ be a category whose objects are unital $C^*$-algebra  and whose arrows are $*$-homomorphisms, then for given $A$ and $B$ belong to $\boldsymbol{u}{\boldsymbol{C}}^{\boldsymbol{*}}$, the map $\phi :B\to A$ which belongs to the $\boldsymbol{u}{\boldsymbol{C}}^{\boldsymbol{*}}\boldsymbol{(}A,B\boldsymbol{)}$\textbf{ }is a nuital $*$-homomorphisms [28]. Now, a question can be asked: whether such $*$-homomorphisms can induce a map ${\underline{\mathrm{\Sigma }}}^A_n\to {\underline{\mathrm{\Sigma }}}^B_{n+1}$? It has been shown that the map $\phi $ induces a map $\widetilde{\phi }:C(B)\to C(A)$ which is called the base map [28]. The map $\widetilde{\phi }$ induces an essential geometric morphism, $\mathrm{\Phi }$, and its inverse image, ${\mathrm{\Phi }}^*$, as follows [28]
\begin{equation} \label{GrindEQ__5_}
\mathrm{\Phi }:\ {\boldsymbol{Sets}}^{C(B)}_{n+1}\to {\boldsymbol{Sets}}^{C(A)}_n
\end{equation}
\begin{equation} \label{GrindEQ__6_}
{\mathrm{\Phi }}^*:\ {\boldsymbol{Sets}}^{{C(A)}^{op}}_n\to {\boldsymbol{Sets}}^{{C(B)}^{op}}_{n+1}
\end{equation}
Therefore, ${\mathrm{\Phi }}^*$ allows us to map the object ${\underline{\mathrm{\Sigma }}}^A_n\in {\boldsymbol{Sets}}^{{C(A)}^{op}}_n$ to the object ${\mathrm{\Phi }}^*({\underline{\mathrm{\Sigma }}}^A_n)\in {\boldsymbol{Sets}}^{{C(B)}^{op}}_{n+1}$. Now, the last step is to define a map ${\mathrm{\Phi }}^*({\underline{\mathrm{\Sigma }}}^A_n)\to {\underline{\mathrm{\Sigma }}}^B_{n+1}$ which for doing that we use the Gelfand duality [28]. Using the duality, it has been shown that there is a natural transformation ${\mathcal{G}}_{\phi }:\ {\mathrm{\Phi }}^*\left({\underline{\mathrm{\Sigma }}}^A_n\right)\to {\underline{\mathrm{\Sigma }}}^B_{n+1}$, in ${\boldsymbol{Sets}}^{{C(A)}^{op}}_n$. Therefore, by combining the two constructed maps ${\mathrm{\Phi }}^*$ and ${\mathcal{G}}_{\phi }$ we obtain the desired map $\left\langle \mathrm{\Phi },{\mathcal{G}}_{\phi }\right\rangle ={\mathcal{G}}_{\phi }\circ {\mathrm{\Phi }}^*:({\underline{\mathrm{\Sigma }}}^A_n)\to {\underline{\mathrm{\Sigma }}}^B_{n+1}$. It has been shown that there exist a bijective correspondence between order-isomorphism $\phi :B\to A$ and the spectral presheaf isomorphisms $\left\langle \mathrm{\Phi },{\mathcal{G}}_{\phi }\right\rangle :({\underline{\mathrm{\Sigma }}}^A_n)\to {\underline{\mathrm{\Sigma }}}^B_{n+1}$ [28].

\noindent Summarizing: given two unital $C^*$-algebras $A$, $B$ belong to unital commutating $C^*$-algebra ($\boldsymbol{uc}{\boldsymbol{C}}^{\boldsymbol{*}}$) with a unital $*$-homomorphism $\phi :B\to A$, then there exist a map
\begin{equation} \label{GrindEQ__7_}
\left\langle \mathrm{\Phi },{\mathcal{G}}_{\phi }\right\rangle ={\mathcal{G}}_{\phi }\circ {\mathrm{\Phi }}^*:{\underline{\mathrm{\Sigma }}}^A\to {\underline{\mathrm{\Sigma }}}^B
\end{equation}
between the respective presheaves going in the opposite direction [28].

\noindent As a complementary description, we pay our attention to the concept of flow. If $A=B$ then an isomorphism $\left\langle \mathrm{\Phi },{\mathcal{G}}_{\phi }\right\rangle \ :{\underline{\mathrm{\Sigma }}}^A\to {\underline{\mathrm{\Sigma }}}^A$ is called an automorphism ($Aut({\underline{\mathrm{\Sigma }}}^A)$) [28] which can used for defining the flow, $F_A$, on spectral presheaf ${\underline{\mathrm{\Sigma }}}^A$ as $F_A:R\to Aut({\underline{\mathrm{\Sigma }}}^A)$. The concept of flow has been used for providing the topos version of the Heisenberg and Schrodinger representation and study the time evolution of the physical quantities in topos quantum theory [28]. But as we mentioned above, the time is nothing more than the change in the physical quantities when we compare the ${(n+1)}^{th}$ -classical snapshot with the $n^{th}$-snapshot, by using the combined map $\left\langle \mathrm{\Phi },{\mathcal{G}}_{\phi }\right\rangle $. If the map $\left\langle \mathrm{\Phi },{\mathcal{G}}_{\phi }\right\rangle $ be an automorphism, it means that, the system is still in $n^{th}$-classical snapshot  and nothings change, and in consequence, the passage of time has no sense.

\section{Interpretation and Predictions}

\noindent Every physical theorizing contains two important achievements in its scope of validity. The first achievement is a mathematical model that describes the existing laboratory data and provides predictions about the future observations. The second is to generalizes the existed interpretation and paradigms or even to introduce the new types of them. In the initial moments of the world creation, it can be assumed that some entities (events) were created. These events are located in a region of space-time (not a point), which in mathematical term, is called locale (Appendix B). There are the causal (energetic correspondence) relationships between some events i.e., they are placed on a world line and form a causal set. This causal set forms a partially ordered set (poset), called timid child. An element of a poset is called node and the causal relation between two elements is called link. Also, it is possible we consider a collection of disconnected graphs. The set is called a gregarious child. Here, we only consider the timid children. It has been shown that, a poset can be considered as a category and is shown by a graph (called Hasse diagram) [27]. It can be assumed that in the early moments of the world creation, there were many relatively independent and separate timid children such that each of them includes a significant number of nodes. Therefore, we are dealing with a many-node system. Now, two questions can be asked: first, what is the mathematical model of the many-node system? and second, what may be its interpretation? In the previous sections of the paper, we mostly introduced the mathematical model and now we will explain this mathematical model i.e., its interpretation.

\noindent First, let us to provide the interpretation of the kinematic part of the many-node theory. It is well known that the survival and stability of a physical system which includes some elements, occurs when its cohesive (binding) energy becomes a negative number. In the other words, it can be imagined that we have brought each element of the physical system from infinity to its current status (locale in mathematical terms (Appendix B)), and after establishing the conditions of equilibrium and stability, the done work (energy) has been stored in the system as binding energy. The stabilized (relaxed) graph is called timid child. Using a suitable Leslie matrix [39], it is possible, one attributes an entropy function to a system including timid and gregarious children and shows that the entropy becomes maximum when the system reaches to its equilibrium and stable situation [33]. Therefore, as it has been explained in section three, if the total structural Hamiltonian of the timid child be $H^L_T$=$\sum^N_{i=1}{(H^L_i+\frac{1}{2}\sum^{M_i}_{j=1}{V^L_{i,j})}}$, the balance between the first and second terms of the Hamiltonian specifies the stability condition of the system. The related observables to the stability are Raman and infrared spectrums. For example, the infrared spectrum of the molecular Hydrogen and the Methane molecule have been reported in the range 3--20 $\mu m$ [40] and 2470-3200 ${Cm}^{-1}$ [41], respectively. Also, in the spectrum of the universe, the cosmic infrared background has been observed at range, $\upsilon =1012-1014$ Hz, and $\lambda =3-300$ $\mu m$ [42]. Therefore, one can consider a graph with some structural variables, such that after optimization, its infrared radiation coincides with the observed spectrum of the universe. Of course, it is possible to obtain more than one optimized structure matching the same observed spectrum from the theoretical point of view. It means that they are homomorphic structures and therefore more experimental data are necessary for specifying which of them is the optimized structure of the final timid child.

\noindent As it is shown in section four, the total nonstructural Hamiltonian of the timid child is $H^{Net}_T=\sum^N_{i=1}{\left(H^{{N_i,m}_k}_{sa}+\frac{1}{2}\sum^{M_i}_{j=1}{V^{N_i}_{l_j}}\right)}$ whose second (interaction) part includes some terms similar to such as $\sum_i{\sum_j{J_{ij}{\hat{S}}_{1i}{\hat{S}}_{2j}}}$ (spin-interaction), $\sum_i{\sum_j{Q_{ij}{\hat{I}}_{1i}{\hat{I}}_{2j}}}$ (charge-interaction) and so on. Bluemer et al., have discussed the high-energy cosmic-ray measurements covering the energy range from the knee to the highest energies [43]. They have referred to the hadronic interaction models such as QGSJET 01 and SIBYLL 2.1 to interpret some measured data and shown that one of the most popular explanations for the origin of the knee is that the spectra at the source exhibit a break and no point sources of charged cosmic rays were found in the knee region [43]. In addition to information extracted from measurements of charged particles, important hints towards the origin of (hadronic) cosmic rays may be derived from observations of high-energy $\gamma $-rays [43]. Therefore, after finding the relaxed structure of the final timid child, logically, one can assume suitable nonstructural Hamiltonian $H^{Net}_T$ for describing the observed high-energy cosmic-ray related to the first moments of the universe creation. Again, there may be different models for explaining the same experimental data. If it is, it means that they are homomorphic models and more new experimental data are necessary for finding the best ones.

\noindent Now, we provide the interpretation of dynamic. Each entity (event or node) occupies a region, instead of a point, in the universe which is called locale in mathematical terms (Appendix B). All of the above descriptions belong to a specific classical snapshot of the final timid children. After creation of a new locale entity (node), if it is added to an available $n^{th}$-timid child with $m^{n^{th}}_k$-representation of abelian self-adjoint operators, i.e., a causal link is created, a new ${(n+1)}^{th}$-timid child is created and in consequence a new relaxed structure and new $m^{{(n+1)}^{th}}_k$-representation, should be considered. In the other words, time is nothing more than sensing the change in physical observables. If no change is sensed, time will be meaningless [32]. Generally, the evolution of system is described by the concept of flow in topos representation of quantum mechanics [27]. In topos many-node theory, the relationship between $n^{th}$-timid child and ${(n+1)}^{th}$-timid child is specified by a combined map $\left\langle \mathrm{\Phi },{\mathcal{G}}_{\phi }\right\rangle $ which was been explained in the section four. Therefore, logically, it is possible we find the experimental data related to the different final relaxed timid child, sort them and finally classify them in the different collections. By comparison between the collections, we find the signature of different combined maps $\left\langle \mathrm{\Phi },{\mathcal{G}}_{\phi }\right\rangle $ i.e., we specify the history related to a collection of a final relaxed timid children.

\noindent Two remained concepts are the critical dimension of a timid child and the hidden concept behind the truth value. As we have mentioned in section three, each timid child has a critical dimension, after which, a physical paradigm shift happens, and for describing the physics in this new scale, we should use new mathematical model, concepts, and interpretation, probably [32]. In the other words, by adding a new link, the cohesive energy, which is a negative number, increases (will be more positive than before) and in consequence the growth process stops. Therefore, it is expected that we find different signatures belong to the different final relaxed timid children in the different scales of cosmic rays and cosmic background radiations related to the first moments of universe creation.

\noindent In quantum mechanics, we deal with an open system which includes observer, observed system, and measurement. We use a relative frequency interpretation or information probability tree [3] for defining the occurrence probability related to a specific experiment i.e., probability is not a derived concept. But in the topos representation of quantum mechanics, the truth values are the fundamental concept and an internal logic in terms of the truth values are used [27]. Therefore, for both open and closed system we deal with a logical interpretation and the probability is a derived concept [27]. For example, when we say the spin value of a 2-spin system (Appendix C) is at range [1.3-2.3] in fact we express a proposition ($A\in \Delta $) about system such that if the system is in the state $\ket{\psi }_1>$ its related truth value $v\left(A\in \Delta ;\ \ket{\psi }_1\right)=1$ and otherwise it is equal to zero. It is convenient to  define the set $T^s:\left\{S\subseteq X\mathrel{\left|\vphantom{S\subseteq X s\in S}\right.\kern-\nulldelimiterspace}s\in S\right\}$ such that $v\left(A\in \Delta ;s\right)=1$ if and only if $A^{-1}(\Delta )\in T^s$ and otherwise it is equal to zero [27]. Here, $X$ is a topological space. Therefore, the truth value of such a statement is denoted as $\left[\left[A^{-1}(\Delta )\in T^s\right]\right]$ [27]. Now, it is possible to define a measure dependent truth object as
$T^{\mu }_r:=\left\{S\subseteq X\mathrel{\left|\vphantom{S\subseteq X \mu (S)\succcurlyeq r}\right.\kern-\nulldelimiterspace}\mu (S)\succcurlyeq r\right\}$ for all $r\in (0,1]$,
where $X$ is a topological space and $\mu $ is the probability measure i.e., $\mu :Sub\left(X\right)\to [0,1]$ [27]. It should be noted that the value $r=0$ is not included. The reason is to avoid obtaining situation in which all propositions are totally true with probability zero [27]. Therefore, by following the three mentioned steps in section three, one can find the approximated operators related to the operator ${\hat{P}}_V$ i.e., ${\delta }^i{\left(\hat{P}\right)}_{V^{\prime}}$ and ${\delta }^o{\left(\hat{P}\right)}_{V^{\prime}}$ ($V^{\prime}\subseteq V)$ and the quantity valued objects $v=[{\delta }^i{\left(\hat{P}\right)}_{V^{\prime}},\ {\delta }^o{\left(\hat{P}\right)}_{V^{\prime}}]$ which shows the probable values of the operator ${\hat{P}}_V$ . Therefore, the probability is a derived concept, internally.

\section{Summary}

\noindent We have assumed that, some entities are created in the first moments of the world creation. They have occupied a region, instead of a point, in space-time (called locale events (nodes), in mathematical terms). It was assumed that there are the energetic causal relationships between all nodes and all of them are placed on the world line based on the general relativity theory. A graph includes $N$ nodes and $N(N-1)/2$ causal relations which is called links. There are two kinds of the collection of graphs. A timid child is a collection of graphs such that all graphs are connected to each other by links and form a whole graph i.e., all nodes are placed on a world line. A gregarious child is a collection of graphs such that they are some graphs which are not connected to any graph i.e., some nodes are placed on the different world lines. We only considered the timid child which is called the many-node system. The many-node system is a partial ordered set and in consequence it is a topos [27]. It was shown that two kinds of general Hamiltonian called structural Hamiltonian ($H^L_T$) and nonstructural Hamiltonian ($H^{Net}_T$) could be attributed to the system. The surveillance and stability of the many-node graph was explained by using $H^L_T$. It was explained that after reaching to the stability, the final graphs would have a critical dimension. Far from the critical dimension, we deal with a new physical system which should be explained by using the new mathematical model and interpretation. Logically, it should be possible we attribute some infrared spectrum to the final relaxed graph which has been found in the cosmic background radiations [41], for finding its detailed model.  Also, it was discussed that the nonstructural Hamiltonian ($H^{Net}_T$) of $n^{th}$-final optimized graph not only includes some abelian self-adjoint operators (called $m^{n^{th}}_k$-representation) but also the interaction terms such as $\sum_i{\sum_j{J_{ij}{\hat{S}}_{1i}{\hat{S}}_{2j}}}$ (spin-interaction), $\sum_i{\sum_j{Q_{ij}{\hat{I}}_{1i}{\hat{I}}_{2j}}}$ (charge-interaction) and so on. It was discussed that, logically, the signature of $H^{Net}_T\ $(i.e., physical observables) might be found in the cosmic rays, especially in high energy and combination regions [43]. Finally, we explained that time has no meaning else the change in the partial physical observables. Therefore, the relation between the state-space of $n^{th}$-final optimized graph and $({n+1)}^{th}$-final optimized graph could be found by using a combined map $\left\langle \mathrm{\Phi },{\mathcal{G}}_{\phi }\right\rangle $. It is the meaning of the dynamic. Finally, we discussed that, the probability is a derived concept by using the concept of the truth value. The detailed model of the explained may-node theory might be found by using the data of cosmic background radiations and cosmic rays in future studies for removing the probable existed homomorphic in mathematical modeling of structural and nonstructural parts of the theory.

\appendix
\section{Topos Quantum Mechanics}

\noindent Based on the introduced literature in topos physical theory [23-26], ${\mathrm{\Sigma }}_{\phi }$ and $R_{\phi }$ stand for state object and quantity object, respectively, both as ground-type symbol in topos (${\tau }_{\phi }$)-representation. Therefore, a physical quantity is shown by an one-to-one map as $A_{\phi }:{\mathrm{\Sigma }}_{\phi }\to R_{\phi }$ and a set of the physical quantities are shown by $F_{{\tau }_{\phi }}({\mathrm{\Sigma }}_{\phi },R_{\phi })$. In addition, if $\tilde{s}\in {\mathrm{\Sigma }}_{\phi }$ and $\widetilde{\mathrm{\Delta }}\in R_{\phi }$ stand for a free variable and a range of physical quantity, respectively, then a formula i.e., a proposition in sub-object classifier ${\mathrm{\Omega }}_{\phi }$ is shown as the map ${\mathrm{\Sigma }}_{\phi }\times R_{\phi }\to {\mathrm{\Omega }}_{\phi }$ and the collection of these formulas are shown by ${\mathrm{\Gamma }}_{\phi }$. In topos quantum mechanics (TQM), the theory is locally considered as a snapshot of locally classical picture at the same time. The collection of the all snapshots contains all necessary information about the physical system and in consequence no information are lost [27, 28]. In TQM, a category of the abelian von Neumann sub-algebra ( $v(H)$) of the bounded operators algebra on the Hilbert space ($B(H)$) is considered i.e., the topos ${\boldsymbol{Sets}}^{{v(H)}^{op}}\cong Sh(v{\left(H\right)}^-)$ is considered where $Sh$ stands for sheaves and $v{\left(H\right)}^-$ is the complete Heyting algebra of the lower sets in $v(H)$ [27, 28]. Here, If $V_{sa}$ be the set of the all self-adjoint operators and the projection operator $\hat{P}$ is assigned to a physical quantity $P$, such that $\hat{P}\notin $ $V_{sa}$, then one can assigns the smallest element in the set $\left\{\hat{R}\in V_{sa}\mathrel{\left|\vphantom{\hat{R}\in V_{sa} \hat{R}{\ge }_s\hat{P};iff\ {\hat{E}}_{\hat{R}}\preccurlyeq {\hat{E}}_{\hat{P}}}\right.\kern-\nulldelimiterspace}\hat{R}{\ge }_s\hat{P};iff\ {\hat{E}}_{\hat{R}}\preccurlyeq {\hat{E}}_{\hat{P}}\right\}$ (denoted by ${\delta }^o{\left(\hat{P}\right)}_V$) to the projection operator $\hat{P}$ in the context $V\subseteq V_{sa}$through outer deseinisation where ${\hat{E}}_{\hat{R}}$  and ${\hat{E}}_{\hat{P}}\ $are the spectrum decomposition of operators $\hat{R}$ and $\hat{P}$, respectively [27, 28]. Similarly, the largest element in the set $\left\{\hat{R}\in V_{sa}\mathrel{\left|\vphantom{\hat{R}\in V_{sa} \hat{R}{\le }_s\hat{P};\ iff\ {\hat{E}}_{\hat{R}}\succcurlyeq {\hat{E}}_{\hat{P}}}\right.\kern-\nulldelimiterspace}\hat{R}{\le }_s\hat{P};\ iff\ {\hat{E}}_{\hat{R}}\succcurlyeq {\hat{E}}_{\hat{P}}\right\}$ (denoted by ${\delta }^i{\left(\hat{P}\right)}_V$) can be assigned to the projection operator $\hat{P}$ in the context $V\subseteq V_{sa}$ through inner deseinisation [27, 28]. Therefore, one can estimate the operator $\hat{P}$ in $V_{sa}$ by the below inequality in each stage $V$ [27, 28]
\begin{equation}
{\delta }^i{\left(\hat{P}\right)}_V\le {\left(\hat{P}\right)}_{V_{sa}}\le {\delta }^o{\left(\hat{P}\right)}_V                                                                                         
\end{equation}
\noindent Since a state can be defined as
${\underline{T}}^{\rho }_V:=\left\{\hat{P}\in V_{sa}\mathrel{\left|\vphantom{\hat{P}\in V_{sa} Tr\left(\hat{P}\rho \right)=1}\right.\kern-\nulldelimiterspace}Tr\left(\hat{P}\rho \right)=1\right\}$,
a truth value of a physical quantity is defined as [27, 28]
\begin{equation}
v{(\delta (\hat{P})\in {\underline{T}}^{\rho })}_{V_{sa}}:=\left\{V\subseteq V_{sa}\mathrel{\left|\vphantom{V\subseteq V_{sa} Tr\left(\rho {\left(\hat{P}\right)}_V\right)=1}\right.\kern-\nulldelimiterspace}Tr\left(\rho {\left(\hat{P}\right)}_V\right)=1\right\}                                                          
\end{equation}
\noindent where, $Tr$ means trace. Using the above definitions, the quantity valued object related to each physical quantity is defined as [27, 28]
\begin{equation}
\delta {\left(\hat{P}\right)}_V\left(\lambda \right):=\left({\delta }^i{\left(\hat{P}\right)}_V\left(\lambda \right),{\delta }^o{\left(\hat{P}\right)}_V(\lambda )\ \right)                                                                                 
\end{equation}
\noindent where, $\lambda \in {\mathrm{\Sigma }}_{\phi }$. As Eq. (A.3) shows, the lowest (highest) probable measured value of the physical quantity $P$ is equal to ${\delta }^i{\left(\hat{P}\right)}_V$ (${\delta }^o{\left(\hat{P}\right)}_V$) when the system is at state $\lambda \in {\mathrm{\Sigma }}_{\phi }$. However, ${\delta }^i{\left(\hat{P}\right)}_V(\lambda )$ and ${\delta }^o{\left(\hat{P}\right)}_V(\lambda )$ are maps : $\downarrow V\to spectrum\ (\hat{P})$ which are called order-reversing ($\mu $)  and order-preserving function ($\nu $), respectively such that ${\delta }^i{\left(\hat{P}\right)}_V$($\lambda )=\lambda ({\delta }^i{\left(\hat{P}\right)}_{V^{\prime}})$ and ${\delta }^o{\left(\hat{P}\right)}_V\left(\lambda \right)=\lambda \left({\delta }^o{\left(\hat{P}\right)}_{V^{\prime}}\right)$ where $V^{\prime}\subseteq V$. Then, the quantity valued object can be written as [28]
\begin{equation}
{\underline{R}}^{\longleftrightarrow }_V:\left\{(\mu ,\nu )\mathrel{\left|\vphantom{(\mu ,\nu ) \mu ,\ \nu :\downarrow V\to R}\right.\kern-\nulldelimiterspace}\mu ,\ \nu :\downarrow V\to R\right\}                                                                                                        
\end{equation}
\section{Topos representation of Space-time}

\noindent In mathematical terms, A frame is a lattice ,$L$ , with all conjunction (meet:$\ \wedge )$ and disjunction (join: $\vee $) which satisfies the below law for every $U$ and $V_i\in L$
\begin{equation}
U\ {\bigvee }_iV_i={\vee }_i\left(U\bigwedge V_i\right)\                                                                                                                    
\end{equation}
\noindent It should be noted that for every two elements $A$ and $B$, the conjunction (meet:$\ \wedge $) means the greatest lower boundary between them and the disjunction (join: $\vee $) means the least upper boundary between them [28]. When, we say objects are in space-time, it means that they occupy some regions of space-time and not a point. Therefore, the explanation of space-time in terms of extended regions is more useful and valuable than points. The concept of locale should be used instead of frames i.e., instead of considering points, it is preferred to consider the open regions as fundamental entities [28]. In the mathematical terms, the category of locales is the opposite category of frames i.e., $Loc={\left(Frames\right)}^{op}$[28, 36]. Therefore, for every topological space $X$, one can construct $Loc(X)$ whose frame includes all open sub-algebra of $X$. Also, for every map $f:X\to Y$ between topological spaces $X$ and $Y$, the associated maps is $Loc\left(f\right):Loc(X)\to Loc(Y)$ [28]. It can be shown that the locale is equivalent to complete Heyting algebra and in consequence the collection of space-time regions is under an intuitive logic rather than the classical Boolean algebra [28].

\noindent Now, a question can be asked: how can one define the local frame $R^4$? It can be shown that if $A$ and $B$ are two frames then the point of their tensor product i.e., $pt(A\otimes B)$ is $\simeq pt(A)\times pt(B)$ [28]. Here, $pt$ stands for points. Therefore, one can now define space-time to be locale $R\otimes R\otimes R\otimes R=:R^4$ constructed by the iteration ${\bigwedge }_i(A\otimes B)=\left({\bigwedge }_iA\right)\times \left({\bigwedge }_iB\right)$ [28]. Also, since for given three frames $A$, $B$, and $C$ and frame homomorphisms $f:A\to C$ and $g:B\to C$ there exist a unique frame homomorphism $h:A\bigotimes B\to C$ such that $f=h\circ i$ and $g=h\circ j$ where $i:A\to A\otimes B$ and $j:B\to A\otimes B$, the object $R^4$ is a coproduct in $sh(v\left(H\right))$ [28]. It should be noted that since the presheaf $pt(R^4)\simeq {\times }_4pt(R)$ the notion of a point can be defined within the locale [28].

\noindent What is about the quantity valued object when we use locale instead of frame? It has been shown that the quantity valued object ${\underline{R}}^{\longleftrightarrow }$ is a sub-locale of $Loc(v(H)\times R)$ , where $R$ is the internal domain including all compact interval $[a,\ b]$ such that for every $a,\ b\in R$  then $a\le b$ [28]. The points of the locale $Loc(v(H)\times R)$ are given by the presheaf ($Loc(v(H)\times R)$) [28].

\section{Example: 2-spin System}

\begin{enumerate}
\item \textbf{ First step:}
\end{enumerate}

\noindent As an example, let us consider 2-spin system such that the $m_k$-class representation only includes spin operator ${\hat{S}}_z$ , with he below spectrum decomposition in his maximal algebra $V={\mathbb{C}}_1{\hat{P}}_1+{\mathbb{C}}_2{\hat{P}}_2+{\mathbb{C}}_3{\hat{P}}_3+{\mathbb{C}}_4{\hat{P}}_4$
\begin{equation}
{\hat{E}}^{{\hat{S}}_z}_V=\hat{0};\ \ if\ \lambda < -2
\end{equation}
\begin{equation}
{\hat{E}}^{{\hat{S}}_z}_V=\hat{P}_4;\ \ if\ -2\le\lambda<0
\end{equation}
\begin{equation}
{\hat{E}}^{{\hat{S}}_z}_V=\hat{P}_3+\hat{P}_2;\ \ if\ 0\le\lambda<2
\end{equation}
\begin{equation}
{\hat{E}}^{{\hat{S}}_z}_V=\widehat{\boldsymbol{1}}=\sum^4_{i=1}{{\hat{P}}}_i;\ \ if\ \lambda>2
\end{equation}
\noindent Where$\ \hat{0}$ is zero operator, ${\hat{P}}_i=\ket{\psi }_i\bra{\psi }_i$, is the projection operators related to the states $\ket{\psi }_1={\left(1,0,0,0\right)}^{Tra}$ ,$\ket{\psi }_2={\left(0,1,0,0\right)}^{Tra}$, $\ket{\psi }_3={\left(0,0,1,0\right)}^{Tra}$, $\ket{\psi }_4={\left(0,0,0,1\right)}^{Tra}$, ${\lambda }_i$ is the eigenvalue of operator ${\hat{P}}_i$, and $Tra$ means transpose. The set of the other non-maximal sub-algebra of $V$ is
\begin{equation}
\downarrow V=\left\{V^{\prime}\in v\left(H\right)\mathrel{\left|\vphantom{V^{\prime}\in v\left(H\right) V^{\prime}\le V}\right.\kern-\nulldelimiterspace}V^{\prime}\le V\right\}=\{V_{{\hat{P}}_i},V_{{\hat{P}}_i,{\hat{P}}_j}\}\                                                                       
\end{equation}
\noindent where, $V_{{\hat{P}}_i}={\mathbb{C}}_1{\hat{P}}_i+{\mathbb{C}}_2\left({\hat{P}}_j+{\hat{P}}_k+{\hat{P}}_l\right)$; $i\neq j,k,l$, and $V_{{\hat{P}}_i,{\hat{P}}_j}={\mathbb{C}}_1{\hat{P}}_i+{\mathbb{C}}_2{\hat{P}}_j+{\mathbb{C}}_3({\hat{P}}_k+{\hat{P}}_l)$; $i\neq j\neq k,l$ and in both cases $i,j,k,l\in \{1,2,3,4\}$.
Therefore, the state-space related to each algebra are as follows
\begin{equation}
{\underline{\mathrm{\Sigma }}}_V=\{{\lambda }_1,{\lambda }_2,{\lambda }_3,{\lambda }_4\}                                                                                                                
\end{equation}
\begin{equation}
{\underline{\mathrm{\Sigma }}}_{V_{{\hat{P}}_i}}=\left\{{\lambda }^{\prime}_i,{\lambda }^{\prime}_j\right\}                                                                                                                           
\end{equation}
\begin{equation}
\ {\underline{\mathrm{\Sigma }}}_{V_{{\hat{P}}_i,{\hat{P}}_j}}=\{{\lambda }^{{\prime}{\prime}}_i,{\lambda }^{{\prime}{\prime}}_j,{\lambda }^{{\prime}{\prime}}_k\}                                                                                                                \end{equation}

\noindent Also, the sieves are
\begin{equation}
S_V=\left\{V,V_{{\hat{P}}_i},V_{{\hat{P}}_i,{\hat{P}}_j}\mathrel{\left|\vphantom{V,V_{{\hat{P}}_i},V_{{\hat{P}}_i,{\hat{P}}_j} i,j\in \{1,2,3,4\}}\right.\kern-\nulldelimiterspace}i,j\in \{1,2,3,4\}\right\}                                                                                        \end{equation}
\begin{equation}
S_{ij}=\left\{V_{{\hat{P}}_i},V_{{\hat{P}}_i,{\hat{P}}_j}\mathrel{\left|\vphantom{V_{{\hat{P}}_i},V_{{\hat{P}}_i,{\hat{P}}_j} i,j\in \{1,2,3,4\}}\right.\kern-\nulldelimiterspace}i,j\in \{1,2,3,4\}\right\}                                                                                          \end{equation}
\begin{equation}
S_i=\left\{V_{{\hat{P}}_i}\mathrel{\left|\vphantom{V_{{\hat{P}}_i} i\in \{1,2,3,4\}}\right.\kern-\nulldelimiterspace}i\in \{1,2,3,4\}\right\}                                                                                                           \end{equation}
\begin{equation}
{\underline{\mathrm{\Omega }}}_V=\left\{{\underline{0}}_{{\underline{\mathrm{\Omega }}}_V},S_V,S_{ij},S_i\mathrel{\left|\vphantom{{\underline{0}}_{{\underline{\mathrm{\Omega }}}_V},S_V,S_{ij},S_i i,j\in \{1,2,3,4\}}\right.\kern-\nulldelimiterspace}i,j\in \{1,2,3,4\}\right\}
 \end{equation}

\noindent where, ${\underline{0}}_{{\underline{\mathrm{\Omega }}}_V}=\left\{\emptyset \right\}$ and ${\underline{\mathrm{\Omega }}}_V$ is called sub-object classifier related to the maximal algebra $V$.

\begin{enumerate}
\item  \textbf{Second Step:}
\end{enumerate}

\noindent Now, as an example let us assume that the system is in state $\ket{\psi }_1={\left(1,0,0,0\right)}^{Tra}$ which is related to the operator ${\hat{P}}_1$ and we consider the context $V_{{\hat{P}}_4}={\mathbb{C}}_1{\hat{P}}_4+{\mathbb{C}}_2({\hat{P}}_1+{\hat{P}}_2+{\hat{P}}_3)$. By assuming ${\mathbb{C}}_1=-2$ and  ${\mathbb{C}}_2=0$ [27], the spectrum decomposition is
\begin{equation}
{\hat{E}}^{{\hat{S}}_z}_V=\hat{0};\ \ if\ \lambda < -2
\end{equation}
\begin{equation}
{\hat{E}}^{{\hat{S}}_z}_V=\hat{P}_4;\ \ if\ -2\le\lambda<0
\end{equation}
\begin{equation}
{\hat{E}}^{{\hat{S}}_z}_V=\hat{P}_1+\hat{P}_2+\hat{P}_3;\ \ if\ 0\le\lambda<2
\end{equation}
\begin{equation}
{\hat{E}}^{{\hat{S}}_z}_V=\widehat{\boldsymbol{1}}=\sum^4_{i=1}{{\hat{P}}}_i;\ \ if\ \lambda>2
\end{equation}
                                                                                       
\noindent

and by assuming ${\mathbb{C}}_1=-2$ and  ${\mathbb{C}}_2=2$ [27], the spectrum decomposition is
\begin{equation}
{\hat{E}}^{{\hat{S}}_z}_V=\hat{0};\ \ if\ \lambda < -2
\end{equation}
\begin{equation}
{\hat{E}}^{{\hat{S}}_z}_V=\hat{P}_4;\ \ if\ -2\le\lambda<0
\end{equation}
\begin{equation}
{\hat{E}}^{{\hat{S}}_z}_V=\hat{0};\ \ if\ 0\le\lambda<2
\end{equation}
\begin{equation}
{\hat{E}}^{{\hat{S}}_z}_V=\widehat{\boldsymbol{1}}=\sum^4_{i=1}{{\hat{P}}}_i;\ \ if\ \lambda>2
\end{equation}

\noindent By comparison (Appendix A) between
${\hat{E}}^{{\hat{S}}_z}_V$ and ${\delta }_1({\hat{E}}^{{\hat{S}}_z}_{V_{{\hat{P}}_4}})$, and also, ${\hat{E}}^{{\hat{S}}_z}_V$ and   ${\delta }_2({\hat{E}}^{{\hat{S}}_z}_{V_{{\hat{P}}_4}})$ it can be concluded that
\begin{equation}
{\delta }_2({\hat{E}}^{{\hat{S}}_z}_{V_{{\hat{P}}_4}}){\le}_s{\left({\hat{S}}_Z\right)}_V{\le}_s{\delta }_1({\hat{E}}^{{\hat{S}}_z}_{V_{{\hat{P}}_4}})                                                                                       
\end{equation}
\noindent Therefore,
${\delta }_2\left({\hat{E}}^{{\hat{S}}_z}_{V_{{\hat{P}}_4}}\right)={\delta }^i{\left({\hat{S}}_Z\right)}_{V_{{\hat{P}}_4}}$and  ${\delta }_1\left({\hat{E}}^{{\hat{S}}_z}_{V_{{\hat{P}}_4}}\right)={\delta }^O{\left({\hat{S}}_Z\right)}_{V_{{\hat{P}}_4}}$.

\begin{enumerate}
\item  \textbf{Third Step:}
\end{enumerate}

\noindent  Now, it is easy to show that
$\left\langle {\psi }_1\mathrel{\left|\vphantom{{\psi }_1 {\delta }^O{\left({\hat{S}}_Z\right)}_{V_{{\hat{P}}_4}} {\psi }_1}\right.\kern-\nulldelimiterspace}{\delta }^O{\left({\hat{S}}_Z\right)}_{V_{{\hat{P}}_4}}\mathrel{\left|\vphantom{{\psi }_1 {\delta }^O{\left({\hat{S}}_Z\right)}_{V_{{\hat{P}}_4}} {\psi }_1}\right.\kern-\nulldelimiterspace}{\psi }_1\right\rangle =2$\textit{ } and $\left\langle {\psi }_1\mathrel{\left|\vphantom{{\psi }_1 {\delta }^i{\left({\hat{S}}_Z\right)}_{V_{{\hat{P}}_4}} {\psi }_1}\right.\kern-\nulldelimiterspace}{\delta }^i{\left({\hat{S}}_Z\right)}_{V_{{\hat{P}}_4}}\mathrel{\left|\vphantom{{\psi }_1 {\delta }^i{\left({\hat{S}}_Z\right)}_{V_{{\hat{P}}_4}} {\psi }_1}\right.\kern-\nulldelimiterspace}{\psi }_1\right\rangle =0$. 
 Then,  the quantity  valued object  $\check{\delta }{(S_z)}_V$ in  $V_{{\hat{P}}_4}$-context when the system is in the state $\ket{\psi }_1$  is at the range [0,2].
\\
\\

\textbf{References:}

\noindent [1] H. Simchi, ``Causality, Uncertainty Principle, and Quantum Spacetime Manifold in Planck Scale ``, J. Mode. Phys. \textbf{13}, 2 (2022) [arXiv: quant-ph/0604212].

\noindent [2] Steven Weinberg, ``Gravity and Cosmology: Principle and Applications of the General Relativity'' (John Wiley \& Sons, 1972).

\noindent [3] H. Simchi, ``Bell's inequality and Probability Trees'', arXiv: quant-ph/0209095.

\noindent [4] J. Ambj{\o}rn, J. Jurkiewicz and R. Loll, ``Quantum gravity, or the art of building spacetime,'' arXiv: hep-th/0604212.

\noindent [5] C. Rovelli, ``Quantum Gravity'' (Cambridge U. Press, New York, 2004).

\noindent [6] E. Bianchi, L. Modesto, C. Rovelli and S. Speziale, ``Graviton propagator in loop quantum gravity,'' Class. Quant. Grav. \textbf{23}, 6989 (2006) [arXiv: gr-qc/0604044].

\noindent [7] D. Oriti, ``The group field theory approach to quantum gravity,'' arXiv: gr-qc/0607032; D. Oriti, ``Group field theory as the microscopic description of the quantum spacetime fluid: a new perspective on the continuum in quantum gravity,'' arXiv: gr-qc/0710.3276.

\noindent [8] G. Volovik, ``The Universe in a Helium Droplet'' (Oxford University Press, 2003).

\noindent [9] M. Visser and S. Weinfurtner, ``Analogue spacetimes: Toy models for 'quantum gravity'','' arXiv: gr-qc/0712.0427.

\noindent  [10] W. G. Unruh, ``Experimental black hole evaporation,'' Phys. Rev. Lett. \textbf{46}, 1351 (1981).

\noindent [11] E. A. Calzetta and B. L. Hu, ``Bose-Novae as Squeezing of Vacuum Fluctuations by Condensate Dynamics,'' arXiv: cond-mat/0207289v3.

\noindent [12] T. Banks, ``TASI lectures on matrix theory,'' arXiv: hep-th/9911068. 25.

\noindent [13] P. Horava, ``Stability of Fermi surfaces and K-theory,'' Phys. Rev. Lett. \textbf{95}, 016405 (2005) [arXiv: hep-th/0503006].

\noindent [14] J. M. Maldacena, ``The large N limit of superconformal field theories and supergravity,'' Adv. Theor. Math. Phys. \textbf{2}, 231 (1998) [Int. J. Theor. Phys. \textbf{38}, 1113 (1999)] [arXiv: hep-th/9711200].

\noindent [15] S. S. Gubser, I. R. Klebanov and A. M. Polyakov, ``Gauge theory correlators from non-critical string theory,'' Phys. Lett. B, \textbf{428}, 105 (1998) [arXiv: hep-th/9802109].

\noindent [16] T. Konopka, F. Markopoulou and L. Smolin, ``Quantum graphity,'' arXiv: hep-th/0611197.

\noindent [17] R. Sorkin, ``Causal Sets: Discrete Gravity (Notes for the Valdivia Summer School)'', proceedings of the Valdivia Summer School, edited by A. Gomberoff and D. Marolf, arXiv: gr-qc/0309009.

\noindent [18] K. Giesel and T. Thiemann, ``Algebraic quantum gravity (AQG). I: Conceptual setup,'' Class. Quant. Grav. \textbf{24}, 2465 (2007) [arXiv: gr-qc/0607099].

\noindent [19] F. Markopoulou, ``Quantum causal histories,'' Class. Quant. Grav. \textbf{17}, 2059 (2000) [arXiv:hep-th/9904009].

\noindent [20] D. P. Rideout and R. D. Sorkin, ``A Classical sequential growth dynamic for causal sets'', Phys. Rev. D \textbf{61}, 024002 (2000) [arXiv: gr-qc/9904062v3].

\noindent [21] Sumati Surya, ``The causal set approach to quantum gravity, arXiv: gr-qc/1903.11544v2.

\noindent [22] C. Rovelli and F. Vidotto, ``Covariant Quantum Gravity: An elementary introduction to Quantum Gravity and Spinfoam Theory'' (Cambridge University Press, 2015).

\noindent [23] A. Doring and C. J. Isham, ``A Topos Foundation for Theories of Physics: I. Formal Languages for Physics'', J. Math. Phys.~\textbf{49}, 053515 (2008) [arXiv: quant-ph/0703060v1].

\noindent [24] A. Doring and C. J. Isham, ``A Topos Foundation for Theories of Physics: II. Daseinisation and Liberation of Quantum Theory'', J. Math. Phys.~\textbf{49}, 053516 (2008) [arXiv: quant-ph/0703062v1].

\noindent [25] A. Doring and C. J. Isham, ``A Topos Foundation for Theories of Physics: III. The Representation of Physical Quantities with arrows'', J. Math. Phys.~\textbf{49}, 053517 (2008) [arXiv: quant-ph/0703064v1].

\noindent [26] A. Doring and C. J. Isham, ``A Topos Foundation for Theories of Physics: IV. Catagories of Systems'', J. Math. Phys.~\textbf{49}, 053518 (2008) [arXiv: quant-ph/0703066v1].

\noindent [27] Cecilia Flori, ``A First Course in Topos Quantum Theory'' (Springer, 2013).

\noindent [28] Cecilia Flori, ``A Second Course in Topos Quantum Theory'' (Springer, 2018).

\noindent [29] J. Krol, ``A model for spacetime: The Role of Interpretation in some Grothendieck Topoi'', Found. Phys. \textbf{36}, 7 (2006).

\noindent [30] S. Wuppuluri and G. Ghiradi, ``Space, Time, and the Limits of the Human Understanding'' (Springer, 2017).

\noindent [31] T. Konopka, F. Markopoulou, and S. Severini, ``Quantum Graphity: a model of emergent locality ``, Phys.Rev. D 77, 104029 (2008) [arXiv:hep-th/0801.0861v2].

\noindent [32] H. Simchi, ``The Concept of Time: A Grand Unified Reaction Platform'', J. Mod. Phys. 13, 2 (2022) [arXiv: physics.gen-ph/1910.07875v2].

\noindent [33] H. Simchi, ``Statistical Representation of Spacetime'', arXiv: gr-qc/2019.12849v2.

\noindent [34] M. Cortes and L. Smolin, ``The Universe as a process of unique events'' Phys. Rev. D 90, 084007 (2014).

\noindent [35] M. Cortes and L. Smolin, ``Quantum Energetic Causal Sets'' Phys. Rev. D 90, 044035 (2014).

\noindent [36] Tore Dahlen ``The Topos-theoretical Approach to Quantum Physics'' (M.S. Thesis, University of Oslo, 2011).

\noindent [37] A. Atland and B. Simons, ``Condensed Matter Field Theory'' (Cambridge, 2010).

\noindent [38] S. Wolters, ``A comparison of two topos-theoretic approach to quantum theory'' arXiv: math-ph/1010.2031

\noindent [39] Paul Cull, Mary Flahive, and Robby Robson, ``Difference Equations'' (Springer, 2005).

\noindent [40] E. Roueff, H. Abgrall, P. Czachorowski, K. Pachucki, M. Puchalski, and J. Komasa, ``The full infrared spectrum of molecular hydrogen'', Astronomy \& Astrophysics, 630, A58 (2019).

\noindent [41] Earle K. Plyler, Eugene D. Tidwell, and Lamdin R. Blaine, ``Infrared Absorption Spectrum of Methane from 2470 to 3200 cm
J. Research National Bureau of Standards-A, Phys. and Chem. 64, 3 (1960).

\noindent [42] R. Hill, K. Masui, D. Scott, Applied Spectroscopy, 72, 663 (2018) [arXiv:astro-ph.Co/1802.03694v2].

\noindent [43] J. Bluemer,~R. Engel,~J.R. Hoerandel, ``Cosmic Rays from the Knee to the Highest Energies'' Prog. Part. Nucl. Phys. 63, 293 (2009) [arXiv: astro-ph.HE/0904.0725v1].

\noindent

\noindent

\end{document}